# Polymer threadings and rigidity dictate the viscoelasticity and nonlinear relaxation dynamics of entangled ring-linear blends and their composites with rigid rod microtubules

Karthik R. Peddireddy, Ryan Clairmont, and Rae M. Robertson-Anderson*

*Department of Physics and Biophysics, University of San Diego, 5998 Alcala Park, San Diego, CA 92110, United States*

*randerson@sandiego.edu

**Abstract**

Mixtures of polymers of varying topologies and stiffnesses display complex emergent rheological properties that often cannot be predicted from their single-component counterparts. For example, entangled blends of ring and linear polymers have been shown to exhibit enhanced shear thinning and viscosity, as well as prolonged relaxation timescales, compared to pure solutions of rings or linear chains. These emergent properties arise in part from the synergistic threading of rings by linear polymers. Topology has also been shown to play an important role in composites of flexible (e.g., DNA) and stiff (e.g., microtubules) polymers, whereby rings promote mixing while linear polymers induce de-mixing and flocculation of stiff polymers, with these topology-dependent interactions giving rise to highly distinct rheological signatures. To shed light on these intriguing phenomena, we use optical tweezers microrheology to measure the linear and nonlinear rheological properties of entangled ring-linear DNA blends and their composites with rigid microtubules. We show that the linear viscoelasticity is primarily dictated by the microtubules at lower frequencies, but their contributions become frozen out at frequencies above the DNA entanglement rate. In the nonlinear regime, we reveal that mechanical response features, such as shear thinning, stress softening and multi-modal relaxation dynamics are mediated by entropic stretching, threading, and flow alignment of entangled DNA, as well as forced de-threading, disentanglement, and clustering. The contributions of each of these mechanisms depend on the strain rate as well as the entanglement density and stiffness of the polymers, leading to non-monotonic rate dependences of mechanical properties that are most pronounced for highly concentrated ring-linear blends rather than DNA-MT composites.

**Introduction**

For decades now, ring polymers have been the subject of great fascination and frustration, not only because of their biological relevance and industrial applications, but also their simple yet unique 'endless' topology. Theories describing the dynamics of entangled linear polymers, based on the well-established tube model introduced by de Gennes and advanced by Doi and Edwards, rely heavily on the free ends of linear chains [1,2]. Reptation, whereby an entangled polymer diffuses 'head-first' in a curvilinear fashion along its confining tube, formed by the surrounding entangling polymers, is predicted to be the dominant relaxation mode for entangled linear chains. Discrepancies between reptation theory and experiment have been rectified by including contour length fluctuations, in which the polymer ends can rapidly relax to accelerate tube disengagement[3-5]; and constraint release, in which an entangled polymer relaxes by the very slow process of the surrounding entangling chains reptating away to release their constraints [6].

Their lack of free ends prohibits straightforward extension of tube models to ring polymers [7,8]. Multiple theories have been proposed to describe the dynamics of entangled rings, which have been modeled as double-folded structures, akin to linear polymers, and amoeba-like lattice animals, similar to branched polymers [7,9-13]. Entangled rings have also been predicted to display glassy dynamics due to threadings between the rings [14,15]. Experiments to test these differing models have been complicated by the near impossibility of producing samples of pure rings without linear polymer 'contaminants' [16,17]. However, this issue had the important positive consequence of revealing emergent dynamics of ring-linear polymer blends [7,16,18,19]. For example, ring-linear blends have been shown to exhibit increased viscosity [16,18,20-23], slower relaxation modes [18,24,25], and more pronounced shear-thinning compared to their pure ring and linear counterparts [18,24,26]. These emergent rheological properties are expected to arise, in part, from the synergistic threading of rings by linear chains [27-31], which inhibits reptation of rings such that they relax primarily via the slower mode of constraint release. Threading events, which have been shown to be most prevalent in blends with comparable concentrations of rings and linear chains[18,28], have also been shown to enhance polymer stretching and alignment along the direction of the strain, compared to their pure ring counterparts, which in turn facilitates shear thinning and elastic Rouse-like retraction [18,20,28].

We previously showed that entangled ring-linear DNA blends with comparable mass fractions of linear and ring DNA ($\Phi_L \approx \Phi_R$) exhibit the strongest shear thinning and largest elastic plateaus that persist over the widest frequency range[18]. This non-monotonic dependence of viscoelastic properties on $\Phi_L$ persists even in the nonlinear regime in which the blends are pushed far from equilibrium by strains and strain rates that are much larger than the characteristic length and time scales of the blends [18]. We also revealed that a subtle change in DNA topology–free (linear) versus connected (rings) ends–had a dramatic effect on the structure and viscoelastic properties of composites of stiff microtubules (MT) polymerized in the presence of entangled linear or ring DNA [32]. Linear DNA facilitated MT polymerization and flocculation, while ring DNA hindered MT polymerization and promoted DNA-MT mixing. These distinct structural properties gave rise to emergent topology-dependent rheology, whereby the plateau modulus $G_N^0$ for ring composites increased monotonically with MT concentration, while $G_N^0$ for linear DNA composites increased dramatically with a small addition of MTs, after which $G_N^0$ surprisingly decreased with further increase in MT concentration [32].

The fascinating emergent rheological properties of equal mass ring-linear DNA blends ($\Phi_L \approx \Phi_R$) and DNA-MT composites beg the questions: What rheological properties will emerge in composites of microtubules and equal mass ring-linear DNA blends? What roles do ring-linear threadings, entanglements, and polymer flexibility play in the hypothesized emergent rheology?

To address these questions, we investigate the linear and nonlinear microrheological properties of entangled ring-linear blends of equal mass fraction ($\Phi_L \approx \Phi_R$) and their composites with microtubules (Fig. 1a). The ring-linear blends are identical in composition, comprising linear and ring DNA of equal contour lengths $L = 39$ μm and mass fractions, but vary in overall DNA concentration ($c_D =0.5$ mg/ml and 0.65 mg/ml). The DNA-MT composite is produced by directly polymerizing $c_T = 0.7$ mg/ml tubulin in the $c_D = 0.5$ mg/ml ring-linear DNA blend. We chose the tubulin concentration $c_T$ such that the corresponding entanglement tube diameter $d_T$ is comparable to that of the 0.65 mg/ml DNA blend (see supplementary materials). In this way, we can directly compare the effects of adding stiff (MTs) or flexible (DNA) polymers to a ring-linear DNA blend.

As shown in Fig. 1a, we denote these three systems by the concentrations of DNA and tubulin in mg/ml: $c_D : c_T = 0.5:0$, 0.65:0, and 0.5:0.7. We use optical tweezers microrheology to systematically compare the viscoelastic properties of the 0.5:0 DNA blend with the 0.65:0 DNA blend and 0.5:0.7 DNA-MT composite in both linear and nonlinear regimes.

We find that while the addition of either DNA or MTs to the 0.5:0 blend increases the magnitudes of the linear viscoelastic moduli and extends the longest relaxation timescale, the frequency dependence is relatively insensitive to increased DNA concentration whereas MTs significantly increase the elastic contribution to the response. Moreover, the effect of MTs on the dynamics become frozen out at higher frequencies in which the dynamics are dominated by the DNA. In the nonlinear regime, we observe complex force response and relaxation dynamics with dependences on strain rate and network composition that are surprisingly distinct from the linear regime. Notably, both 0.5 mg/ml systems display similar nonlinear rheological features, including strain softening and reduced shear thinning and relaxation timescales, while 0.65:0 exhibits enhanced shear thinning and unique non-monotonic rate dependences of rheological features.

**Methods**

**DNA**: We prepare double-stranded, 115 kilobasepair (kbp) DNA by replication of cloned bacterial artificial chromosomes (BACs) in *Escherichia coli*, followed by extraction, purification, concentration and resuspension in TE10 buffer (10 mM Tris-HCl (pH 8), 1 mM EDTA, 10 mM NaCl) using custom-designed protocols described and validated previously [33,34]. We use gel electrophoresis image analysis to quantify the DNA concentration and mass fractions of relaxed circular (ring) and linear topologies. Using Life Technologies E-Gel Imager and Gel Quant Express software we determine $c \simeq 0.8$ mg/ml and $\Phi_L \simeq \Phi_R$. For experiments, we use two different dilutions of this stock: $c_{D,\downarrow} = 0.5$ mg/ml and $c_{D,\uparrow} = 0.65$ mg/ml. More details about the physical properties of the DNA are provided in Supplementary Materials.

**Microtubules (MT)**: Unlabeled and rhodamine-labeled porcine brain tubulin (Cytoskeleton, Inc; T240, TL590M) are stored at -80°C in single-use aliquots containing 5 mg/ml tubulin dimers with an unlabeled:labeled ratio of 9:1 in PEM100 buffer (100 mM PIPES (pH 6.8), 2 mM $MgCl_2$, 2mM EGTA). As previously described [32], to form DNA-MT composites, we add a concentration of $c_T = 0.7$ mg/ml tubulin dimers to the $c_D = 0.5$ mg/ml DNA blend, followed by 2 mM GTP and 10 µM Taxol to polymerize and stabilize the MTs. The formed MTs are hollow rods with a diameter $D \simeq 25$ nm comprising 13 tubulin dimers per ring[35].

**Sample preparation**: We perform measurements on three different systems, which we denote by the ratio of the concentrations of DNA and tubulin in mg/ml: $c_D:c_T = 0.5:0$, 0.65:0, and 0.5:0.7. For microrheology experiments, we add a trace amount of 4.5 µm polystyrene microspheres (probes), coated with Alexa-488 BSA to prevent DNA adsorption and enable fluorescence imaging, and an oxygen scavenging system (45 µg/mL glucose, 43 µg/mL glucose oxidase, 7 µg/mL catalase, and 5 µg/mL β-mercaptoethanol) to inhibit photobleaching. Sample chambers (20×3×0.1 $mm^3$) are made with a microscope glass slide and coverslip, each coated with BSA to prevent adsorption of DNA, MTs and beads, and separated by a layer of double-sided tape. All samples are mixed slowly and thoroughly, using wide-bore pipette tips to prevent shearing, then introduced into sample chambers through capillary action and hermetically sealed with epoxy. For the DNA-MT composite (0.5:0.7), tubulin dimers are added immediately before flowing into the chamber, and the sample chamber is incubated at 37°C for 2 hours, resulting in repeatable and reliable tubulin polymerization in the DNA blend [32].

**Microrheology**: We use optical tweezers microrheology to determine the linear and nonlinear rheological properties of the three systems (Fig 1) [36,37]. The optical trap, built around an Olympus IX71 epifluorescence microscope, is formed from a 1064 nm Nd:YAG fiber laser (Manlight) focused with a 60× 1.4 NA objective (Olympus). Forces exerted by the polymer networks on the trapped beads are determined by recording the laser beam deflections via a position sensing detector (Pacific Silicon Sensors) at 20 kHz. The trap is calibrated for force measurement using Stokes drag method [38-40]. All microrheological data is recorded at 20 kHz, and at least 15 trials are conducted, each with a new microsphere in an unperturbed location. Presented data is an average of all trials and error bars represent standard error.

*Linear microrheology:* We determine linear viscoelastic properties from the thermal fluctuations of trapped microspheres, measured by recording the associated laser deflections for 200 seconds (Fig 1e). We extract the elastic modulus $G'(\omega)$ and viscous modulus $G''(\omega)$, from the thermal fluctuations using the generalized Stokes-Einstein relation (GSER) as described in ref [41]. In brief, we compute the normalized mean-squared displacements ($\pi(\tau) = \langle r^2(\tau) \rangle / 2 \langle r^2 \rangle$) of the thermal forces, averaged over all trials, which we convert into the Fourier domain via:

$$-\omega^2\, \pi(\omega) = (1 - e^{-i\omega\tau_1})\frac{\pi(\tau_1)}{\tau_1} + \dot{\pi}_\infty e^{-i\omega t_N} + \sum_{k=2}^{N}\left(\frac{\pi_k - \pi_{k-1}}{\tau_k - \tau_{k-1}}\right)\left(e^{-i\omega\tau_{k-1}} - e^{-i\omega\tau_k}\right),$$

where $\tau$, 1 and $N$ represent the lag time and the first and last point of the oversampled $\pi(\tau)$. $\dot{\pi}_\infty$ is the extrapolated slope of $\pi(\tau)$ at infinity. Oversampling is done using the MATLAB function PCHIP. $\pi(\omega)$ is related to viscoelastic moduli via:

$$G^*(\omega) = G'(\omega) + iG''(\omega) = \left(\frac{k}{6\pi R}\right)\left(\frac{1}{i\omega\pi(\omega)} - 1\right),$$

where $R$ and $k$ represent the microsphere radius and trap stiffness. From $G'(\omega)$ and $G''(\omega)$, we compute the complex viscosity $\eta^*(\omega) = [(G'(\omega))^2 + (G''(\omega))^2]^{1/2}/\omega$ and loss tangent $\tan\delta = G''(\omega)/G'(\omega)$.

*Nonlinear microrheology:* We perform nonlinear microrheology measurements by displacing a trapped microsphere embedded in the sample through a distance $x = 15\ \mu m$ at logarithmically spaced speeds of $v = 10 - 200\ \mu m/s$ using a piezoelectric nanopositioning stage (Mad City Laboratories) to move the sample relative to the microsphere (Fig 1f). We convert the distance to strain via $\gamma = x/2R = 3.4$, and convert speed to strain rate via $\dot{\gamma} = 3v/\sqrt{2}R\ (= 9.4 - 189\ s^{-1})$ [42].

**Results and Discussion**

*Linear Viscoelasticity (LVE)*

We first examine the linear viscoelasticity (LVE) of the three systems, $c_D : c_T = $ 0.5:0, 0.65:0, 0.5:0.7, by extracting the frequency-dependent elastic and viscous moduli, $G'(\omega)$ and $G''(\omega)$, from the thermal fluctuations of trapped beads (see Methods, Figs. 1b,e and 2). As shown in Figure 2a, both DNA blends (i.e., 0.5:0, 0.65:0) display similar frequency dependence of $G'(\omega)$ and $G''(\omega)$, indicative of modestly entangled polymers, with an entanglement regime, where $G'(\omega) > G''(\omega)$, flanked by higher and lower frequency regimes in which dissipative dynamics dominate (i.e. $G''(\omega) > G'(\omega)$). For reference, the low (slow) and high (fast) frequencies at which $G'(\omega)$ and $G''(\omega)$ crossover, $\omega_{c,s}$ and $\omega_{c,f}$, indicate the slowest ($\omega_{c,s}$) and fastest ($\omega_{c,f}$) relaxation rates of the system, which correspond to corresponding relaxation timescales $\tau_c = 2\pi/\omega_c$. While the frequency dependence of $G'(\omega)$ and $G''(\omega)$ are similar for both DNA blends, the magnitude of the moduli and the frequency range over which $G''(\omega) > G'(\omega)$ are both larger for the higher concentration blend (0.65:0), as one may expect given the increased entanglement density [1,2,43-45].

According to Doi-Edwards theory for linear entangled polymers [2], $\tau_{c,f}$ corresponds to the entanglement time $\tau_e$, i.e., the time it takes for a confined polymer to 'feel' its tube confinement. The corresponding frequency values for 0.5:0 and 0.65:0 are $\omega_{c,f,0.5} \simeq 5.6$ rad/s and $\omega_{c,f,0.65} \simeq 16$ rad/s, respectively, such that $\tau_{c,f,0.5} \simeq 1.1$ s and $\tau_{c,f,0.65} \simeq 0.39$ s. In comparison, the theoretically predicted entanglement times for linear DNA at $c = 0.5$ mg/ml and 0.65 mg/ml are $\tau_{e,0.5} \simeq 0.17$ s and $\tau_{e,0.65} \simeq 0.10$ s. Both experimental and theoretical timescales are ~2× shorter for 0.65:0 compared to 0.5:0, in line with the Doi-Edwards scaling relations $\tau_e \sim a^4 \sim L_e^2 \sim c^{-2}$. Namely, increasing polymer concentration reduces the length between entanglements $L_e$ and corresponding tube diameter $a$ and entanglement time $\tau_e$. While the scaling of $\tau_e$ with $c$ for ring-linear blends qualitatively aligns with predictions for linear polymers, the magnitudes of $\tau_{c,f,0.5}$ and $\tau_{c,f,0.65}$ are ~5.2× higher, suggesting a lower entanglement density (i.e. larger $L_e$) than their pure linear polymer counterparts. This result aligns with the general concept that ring polymers are less effective at forming entanglements than linear chains due to their lack of free ends [7,9,23,46-48].

The low (slow) crossover frequency $\omega_{c,s}$ for entangled linear polymers is predicted to correspond to the disengagement rate $\omega_D = 2\pi/\tau_D$, where the disengagement time $\tau_D$ is the time for a deformed polymer to reptate out of its entanglement tube. The predicted $\tau_D$ values for $c = 0.5$ mg/ml and 0.65 mg/ml linear DNA are $\tau_{D,0.5} \simeq 7.4$ s and $\tau_{D,0.65} \simeq 10$ s, respectively. The corresponding slow timescales for 0.5:0 and 0.65:0 blends, which we compute from measured crossover frequencies $\omega_{c,s,0.5} \simeq 0.16$ rad/s and $\omega_{c,f,0.65} \lesssim 0.1$ rad/s, are $\tau_{c,s,0.5} \simeq 39$ s and $\tau_{c,s,0.65} \gtrsim 63$ s. Similar to the fast timescales, the concentration dependence of $\tau_{c,s}$ aligns with theoretically predicted scaling $\tau_D \sim L_e^{-1} \sim c$, but the magnitudes differ substantially. In this case, the experimentally measured relaxation times are ~5.8× longer than the predicted timescales for entangled linear chains. This result is rather surprising given that our measured fast timescales indicate that $L_e$ is larger in the ring-linear blends compared to their linear counterparts. Specifically, given the relations $L_e \sim (\tau_{c,f})^{1/2}$ and $L_e \sim \tau_D^{-1}$ our measured $\tau_{c,f}$ values imply that $L_e$ for the blends is ~2.3× larger than for pure linear chains, such that $\tau_{c,s}$ should be ~2.3× lower.

We can understand this substantially slower relaxation in blends by recalling that threadings between the ring and linear polymers have been shown to play a principal role in the relaxation dynamics of ring-linear blends [18,20,23,24,27-31,49-51]. Relaxation of threaded polymers has been shown to be much slower than reptation, as it relies primarily on constraint release [24,27,28,31,52,53]. Our slow relaxation timescales, which are an order of magnitude slower than theoretically predicted for linear polymers, assuming the increased entanglement length that our $\tau_{c,f}$ values indicate (i.e., ~2.2 ∗ 5.8), are strong evidence that threading plays a dominate role in the relaxation dynamics of both DNA blends.

In Fig. 2b, we evaluate the effect of adding microtubules to the 0.5:0 blend (i.e., $c_D : c_T = 0.5:0.7$), rather than more DNA (as in Fig 2a). Similar to increasing DNA concentration, we observe that the magnitudes of $G'(\omega)$, $G''(\omega)$, and $\omega_{c,f}$ increase when MTs are added to 0.5:0. However, unlike the 0.65:0 blend, the frequency-dependence of $G'(\omega)$ for 0.5:0.7 is weaker than the 0.5:0 system. While both DNA blends roughly exhibit $G'(\omega) \sim \omega^{0.3}$ scaling (Fig 2a), the 0.5:0.7 composite displays a much weaker scaling of $G'(\omega) \sim \omega^{0.1}$, akin to a rubbery plateau. We previously observed similarly weakened $G'(\omega) \sim \omega^{0.1}$ scaling for DNA-MT composites comprising either pure ring or pure linear DNA, indicating that this scaling is relatively insensitive to variations in entanglement density and threading propensity of the DNA [32]. Interestingly, this scaling is weaker than the reported $\omega^{0.17}$ scaling for pure entangled MTs at ~40% higher concentration [54]. Taken together, our results suggest that synergistic interactions between the flexible DNA and stiff MTs enhance the elasticity of the composite beyond that of their monodisperse counterparts [32]. Similar enhanced stiffness and elasticity has been reported for other composites of stiff and flexible polymers [55,56].

The viscous modulus $G''(\omega)$ is likewise larger and exhibits much weaker frequency dependence for the 0.5:0.7 composite compared to 0.5:0 over two decades of $\omega$. This minimal $\omega$-dependence is similar to that reported for pure MTs at 1 mg/ml over a similar frequency range [54]. However, for $\omega > \omega_{c,f,0.5}$, $G''(\omega)$ for both systems, with and without MTs, collapse to a single curve with increased $\omega$-dependence ($G''(\omega) \sim \omega^{3/4}$). This collapse, along with similar high-$\omega$ scaling for 0.65:0 (Fig 2a) and minimal $\omega$ dependence for pure MT solutions [54], suggests that the DNA dynamics dominate the viscoelasticity for timescales shorter than the entanglement time $\tau_e$ of the DNA. At these short timescales (high frequencies), the dynamics are largely dissipative as the DNA polymers do not feel their entanglement constraints. Moreover, the prohibitively slow relaxation timescales for MTs prevent appreciable contribution to the dynamics [57].

We more closely examine this high frequency regime by comparing the fast crossover frequencies $\omega_{c,f}$ for 0.5:0 and 0.5:0.7, which we measure to be $\omega_{c,f,0.5} \simeq 5.6$ rad/s and $\omega_{c,f,0.5:0.7} \simeq 31$ rad/s, respectively. As such, while the viscous contribution to the dynamics of both systems is identical for $\omega \gtrsim 10$ rad/s, as discussed above, the onset of entanglement dynamics and elastic contributions appears to occur at ~5.6× shorter timescales for 0.5:0.7 compared to 0.5:0. Namely, the fast relaxation timescales, which we understand to be indicative of the entanglement time $\tau_e$, are $\tau_{c,f,0.5:0} \simeq 1.1$ s and $\tau_{c,f,0.5:0.7} \simeq 0.20$ s for the DNA blend and DNA-MT composite, respectively. $\tau_{c,f,0.5:0.7}$ is also nearly ~2-fold shorter than that for the 0.65:0 blend. This result suggests that the presence of rigid MTs increases the effective entanglement density of the DNA blend more so than the presence of more DNA. Finally, we note that the DNA-MT composite does not exhibit a low-$\omega$ crossover suggesting that disengagement processes are slowed substantially by the presence of rigid MTs which, themselves, have relaxation timescales that can span minutes to hours [54,57].

To further elucidate the differences in the viscoelastic properties between the three networks, we compare their respective complex viscosities $\eta^*(\omega)$ (Fig. 2c). To understand the trends shown in Fig 2c, we first note that entangled polymers typically exhibit shear thinning, in which systems that obey the Cox-Merz rule follow $\eta^*(\omega) \sim \omega^{-\alpha}$ scaling with $0 \leq \alpha \leq 1$ [2,58-60]. This behavior is expected to be a result of flow alignment of the polymers and reaches a maximum of $\alpha \simeq 1$ for highly entangled polymer solutions and gels [2,20,26,38,60-62]. Entangled DNA solutions and blends have been reported to display weaker thinning of $\alpha \simeq 0.4 - 0.7$, largely insensitive to DNA concentration, but exhibiting maximal thinning for equal mass ring-linear blends and weakest thinning for pure rings [18,20,26,32,60]. Shear thinning is clearly evident for all three systems, with both DNA systems (0.5:0, 0.65:0) following similar scaling of $\alpha \simeq 0.65$ and the 0.5:0.7 composite obeying $\alpha \simeq 1$ for all but the highest frequencies in which the thinning of all systems is slightly weaker. This onset of weakened high-$\omega$ thinning, which indicates that the polymers do not have time to align with flow, appears to be $\sim \omega_{c,f}$. As described above, entanglements, which mediate viscosity thinning, have a negligible role on dynamics for $\omega > \omega_{c,f}$.

Comparing the magnitudes of $\eta^*(\omega)$ across the networks ($\eta^*_{0.65}$, $\eta^*_{0.5}$, $\eta^*_{0.5:0.7}$) we observe that $\eta^*_{0.65}$ is ~2.3-fold larger than $\eta^*_{0.5}$ across the entire frequency range, similar to the corresponding $G'(\omega)$ and $G''(\omega)$ curves, and as expected given the comparatively higher concentration and thus shorter entanglement length $L_e$ of 0.65:0. Intriguingly, we find that for low frequencies ($\omega \lesssim 3$ rad/s), the DNA-MT viscosity ($\eta^*_{0.5:0.7}$) is ~2× and ~4× larger than $\eta^*_{0.65}$ and $\eta^*_{0.5}$, respectively, indicating substantially increased resistance to flow; however, $\eta^*_{0.5:0.7}$ drops below $\eta^*_{0.65}$ at higher frequencies. In fact, at the highest frequencies $\eta^*_{0.5:0.7}$ is comparable to $\eta^*_{0.5}$, indicating that the microtubules do not contribute significantly to the viscoelastic response in the high-$\omega$ limit. We can understand this short-timescale phenomena as arising from the coupled effects of the DNA not having enough time to feel their tube constraints formed by the entangling microtubules ($t < \tau_e$) and the microtubules not having time to appreciably relax. In other words, the contribution of the microtubules to the viscoelastic response becomes increasingly 'frozen out' at sufficiently high frequencies.

The results described above and in Fig 2a-c, indicate that the different networks undergo varying degrees of dissipation versus storage, which we quantify by the loss tangent $\tan\delta = G''(\omega)/G'(\omega)$ (Fig 2d). For reference, elastic processes dominate the response when $\tan\delta < 1$, with lower values indicating increasing elasticity; while $\tan\delta > 1$ indicates that dissipation dominates the response with higher values indicating relatively more dissipation. The frequencies at which $\tan\delta = 1$ demarcate the crossover frequencies, $\omega_{c,s}$ and $\omega_{c,f}$, discussed above. As shown, the DNA-MT composite is substantially more elastic than the DNA blends, which show similar $\tan\delta$ values between them (Fig 2c,f), but the effect of MTs is more subdued at higher frequencies where $\tan\delta > 1$. The frequencies at which $\tan\delta = 1$, which align with the values we report above for $\omega_{c,f,0.5}, \omega_{c,f,0.65}$, and $\omega_{c,f,0.5:0.7}$, show that adding either DNA or MTs to the 0.5:0 blend extends the timescale of the elastic-like entanglement regime (i.e., where $\tan\delta < 1$). However, adding stiff MTs extends the entanglement regime over longer timescales than adding more flexible DNA ($\omega_{c,f,0.5:0.7} > \omega_{c,f,0.65}, \omega_{c,s,0.5:0.7} < \omega_{c,s,0.65}$), and likewise suppresses the dissipative contributions to the entanglement regime by ~2-fold.

Our collective linear microrheology results reveal that the addition of either DNA or MTs to the 0.5:0 DNA blend, increases the magnitudes of the viscoelastic moduli ($G'(\omega)$, $G''(\omega)$, $\eta^*(\omega)$) and extends the entanglement regime by increasing $\omega_{c,f}$ and decreasing $\omega_{c,s}$. However, increased DNA has minimal effect on the frequency dependence of the moduli, suggesting that similar mechanisms dictate the dynamics in both DNA blends. In contrast, MTs substantially weaken the frequency dependence of $G'(\omega)$, increase the degree of viscosity thinning, and decrease the loss tangent, all indicators of increased elasticity and rigidity. Notably, these effects that emerge in the 0.5:0.7 composite are frozen out at higher frequencies that are faster than the DNA entanglement rate ($\tau_e^{-1}$) and the fastest relaxation rate of the MTs.

*Nonlinear Viscoelastic Response*

We next seek to determine the extent to which LVE features and mechanisms described above are preserved in the nonlinear regime in which we subject the polymers to large strains $\gamma$ and fast strain rates $\dot{\gamma}$ (Figs. 1c, 1f, 3, and S1). As described in Methods and Fig 1, to impart nonlinear strains, we drive the same optically trapped microspheres that we use to measure linear viscoelasticity through a strain of $\gamma = 3$ at logarithmically spaced rates of $\dot{\gamma} = 9.4 - 189 \text{ s}^{-1}$ (Fig. 3) while measuring the force the network exerts on the moving probe. To understand the force response curves shown in Figs. 3a-b and S1, we first recall that for a purely elastic material, $F(\gamma, \dot{\gamma}) \sim \gamma \sim \dot{\gamma}^0$ such that $F$ should increase linearly with $\gamma$ with a $\dot{\gamma}$-independent slope that indicates the network stiffness $K = dF/d\gamma$. Conversely, for a purely viscous solution, $F(\gamma, \dot{\gamma}) \sim \gamma^0 \sim \dot{\gamma}$, such that $F$ should immediately rise to a maximum $\gamma$-independent plateau with a magnitude that scales linearly with $\dot{\gamma}$. Fig 3 shows that all three networks generally exhibit an initial steep rise in force before rolling over to a 'softer' regime with a roughly constant positive slope that is weaker than that of the initial rise. Further, the maximum force $F_{max}$ that each network exerts during strain increases monotonically with increasing $\dot{\gamma}$ (Fig. 3e inset). However, as we describe below, the magnitudes and slopes of $F(\gamma, \dot{\gamma})$ in the initial 'stiff' and subsequent 'soft' regimes, and the time and strain distance at which the rollover to the soft regime occurs ($t_{soft}$ and $d_{soft}$), all depend on the network composition and strain rate.

Comparing the nonlinear response of the two DNA blends (0.5:0, 0.65:0), we observe that, for a slow strain rate of $\dot{\gamma} = 19$ s$^{-1}$, the shape of the curves for both networks are similar, but the 0.65:0 exhibits a substantially stronger resistive force for the entire strain $\langle F_{0.65}\rangle \simeq 1.5 \langle F_{0.5}\rangle$, as we may expect given the increased entanglement density and linear viscoelasticity. However, for a 10-fold faster strain of $\dot{\gamma} = 189$ s$^{-1}$ the $\gamma$-dependence of $F(\gamma)$ differs substantially between the networks. Namely, 0.65:0 exhibits a more solid-like response with a weaker initial rise in $F(\gamma)$ and steeper 'soft' regime compared to 0.5:0, suggesting that different mechanisms drive the nonlinear response of 0.65:0 versus 0.5:0. Moreover, the initial force response of 0.65:0 is surprisingly smaller than 0.5:0 and only becomes larger at $\gamma \geq 1.7$. To better quantify and examine this intriguing behavior and its dependence on strain rate, we plot the difference of the force curves for 0.65:0 and 0.5:0, $F_{0.65} - F_{0.5}$, as a function of $\gamma$ for all strain rates (Fig 3c). Negative values indicate that the force response of the 0.65:0 blend ($F_{0.65}$) is (counterintuitively) lower than that of 0.5:0. As shown, this crossover behavior also manifests for $\dot{\gamma} = 90$ s$^{-1}$ and 42 s$^{-1}$ with the magnitude of the difference and the strain over which it persists decreasing with decreasing $\dot{\gamma}$.

To understand this intriguing behavior, we turn to previous optical tweezers microrheology studies that reported evidence of polymer build-up at the leading edge of the probe as well as strain-induced de-threading and alignment with flow[18,63]. In these studies, higher concentration ring-linear DNA blends displayed increased flow alignment, facilitated by threading events, which manifested as more pronounced nonlinear shear thinning. At lower concentrations, in which blends had fewer entanglements and threading events, the moving probe could de-thread and disentangle polymers, allowing them to preferentially accumulate at the leading edge, rather than aligning with the strain. This build-up led to reduced shear-thinning. Our results are consistent with these previous findings and elucidate the force response that these different mechanisms give rise to. At higher strain rates, DNA polymers in the 0.65:0 blend remain constrained by entanglements and threadings such that they are stretched along the strain path, leading to an elastic-like force response ($F \sim \gamma$) over much of the strain, with lower initial $F$ values compared to 0.5:0 due to the lack of polymer build-up and the minimal stretching and alignment that occurs at these shorter lengthscales. At lower strain rates, the polymers have more time to disentangle and/or de-thread in response to the moving probe, allowing polymers to more easily build-up in front of the probe, such that the strain dependence of the force response more closely resembles that of the lower concentration DNA blend.

We observe very different effects when adding MTs to 0.5:0 compared to adding more DNA. As shown in Fig 3b, $F_{0.5:0.7}$ is significantly larger than $F_{0.5}$ over the entire strain for both low and high strain rates, but the shapes of the curves are similar, suggesting similar mechanisms driving the nonlinear response. This effect can be seen more clearly in Fig 3d that shows that $F_{0.5:0.7} - F_{0.5}$ curves are positive for all $\gamma$ and $\dot{\gamma}$ and larger in magnitude than the corresponding difference curves for the two DNA blends (Fig 3c). The steep initial rise in force for both systems, followed by a weak strain dependence, is suggestive of polymer build-up at the leading edge of the probe, as we describe above. However, at faster strain rates, the 0.5:0.7 composite exhibits stronger strain dependence, indicative of increased elasticity, as we see in the linear regime. The rigid and slow MTs provide substantially more resistance to strain than more DNA as they are unable to affinely reorient and stretch on the timescale of the nonlinear strains, with rates that are nearly all faster than the frequency range we examine in LVE measurements ($\omega = 2\pi\dot{\gamma} \simeq 60 - 1190$ rad/s). Further, the rigid MT scaffold prevents the entangling DNA from stretching and aligning with the flow, instead facilitating strain-induced disentanglement and build-up at the leading edge.

To more quantitatively compare the $\dot{\gamma}$-dependence of the nonlinear force response of all three networks, we plot the corresponding maximum force $F_{max}$ reached during the strain and effective nonlinear viscosity $\eta_{en} = F_{max}/\dot{\gamma}$, both as functions of $\dot{\gamma}$ (Fig 3e). Corroborating the trends shown in Fig 3a,b, we find that $F_{max}$ for the DNA-MT composite (0.5:0.7) is ~3-fold larger than $F_{max}$ for 0.5:0 across all strain rates, with both data sets increasing monotonically with $\dot{\gamma}$ with similar power-law scaling of $F_{max}\sim\dot{\gamma}^{0.45}$. $F_{max}$ for 0.65:0 falls in between $F_{max,0.5:0.7}$ and $F_{max,0.5:0}$ for all $\dot{\gamma}$ but displays weaker increase with $\dot{\gamma}$ (~0.34).

Sublinear scaling of $F_{max}$ with $\dot{\gamma}$ is indicative of nonlinear shear-thinning, which we evaluate by considering the power-law scaling of $\eta_{en}\sim\dot{\gamma}^{-\beta}$ where $\beta$ is the nonlinear analog to $\alpha$ (Fig 2e). We find $\beta$ values of $\beta_{0.5:0}\simeq 0.56\pm 0.01$, $\beta_{0.65:0}\simeq 0.66\pm 0.02$, and $\beta_{0.5:0.7}\simeq 0.53\pm 0.04$ for 0.5:0, 0.65:0 and 0.5:0.7, respectively. We first note that $\beta_{0.5:0}\simeq\beta_{0.5:0.7}$ with values that are lower than their corresponding linear regime thinning exponents, suggesting, as described above, that similar mechanisms drive the nonlinear stress response of both 0.5 systems. Weaker thinning suggests reduced flow alignment which likely arises from the faster strain rates which facilitate the disentanglement and de-threading of DNA and hinder the ability of MTs to reorient and align with the strain, both effects suppressing shear-thinning and promoting build-up. At first glance, these results appear at odds with our LVE measurements in which the 0.5:0.7 composite displays much stronger thinning than 0.5:0. However, as we describe in the preceding section, this enhanced thinning of the DNA-MT composite becomes weaker with increasing $\omega$, eventually overlapping with that of 0.5:0 at frequencies comparable to the lowest nonlinear strain rate where MT contributions are largely frozen out.

We now return to the early-time strain dependence for the different networks, most clearly seen in Fig 3a,c, that shows that the time, $t_{soft}$, and lengthscale, $d_{soft}=vt_{soft}$, at which $F(\gamma,\dot{\gamma})$ rolls over from the initial 'stiff' regime to the subsequent 'soft' regime are strongly dependent on $\dot{\gamma}$. As shown in Fig. 3f, $t_{soft}$ and $d_{soft}$ are both largely insensitive to network composition, but $t_{soft}$ exhibits a surprising non-monotonic dependence on $\dot{\gamma}$, reaching a maximum at $\dot{\gamma}=42\text{ s}^{-1}$. Previous differential dynamic microscopy measurements on similar systems[64] reported that flow alignment and entropic stretching of DNA followed an analogous non-monotonic $\dot{\gamma}$ dependence that peaked at $\dot{\gamma}=42\text{ s}^{-1}$. We can understand this non-monotonic trend by considering that rates below this 'resonant' rate, the polymers have time to relax and disentangle (dissipative processes) to facilitate transitioning to a softer, more fluid-like regime. Rates above this resonance are too fast for the polymers to effectively stretch and align with the flow and instead the moving probe forces disentanglement and de-threading.

By scaling $t_{soft}$ by the strain speed $v$, we show that the softening lengthscale $d_{soft}$ undergoes an initial steep increase with increasing $\dot{\gamma}$ which largely plateaus for $\dot{\gamma}\geq 42\text{ s}^{-1}$. The average $d_{soft}$ values in this plateau regime are $d_{soft}=8.1\pm 2.5$ µm, $7.2\pm 1.3$ µm, and $6.6\pm 1.3$ µm for 0.5:0, 0.65:0, and 0.5:0.7, respectively. Notably, these lengthscales align with theoretically predicted values for the entanglement lengths $L_{e,0.5:0}\simeq 10$ µm and $L_{e,0.65:0}\simeq 8$ µm for the DNA solutions. This agreement indicates that softening to a more dissipative regime occurs after the polymers have been entropically stretched to their maximum length set by the confining entanglements. Beyond $L_e$ the polymers are prohibited from stretching further and thus undergo dissipative processes such as disentanglement, reptation, and constraint release.

While $d_{soft}$ and $t_{soft}$ are largely insensitive to network composition, there are clear differences between the slopes and curvatures of the force curves for each system in both the 'stiff' and 'soft' regimes. To quantify these variations, we compute an effective differential modulus or stiffness, $K(\gamma)=dF/d\gamma$, where $K=0$ for purely viscous system (Fig 4). As shown in Fig 4a-c, for all systems and strain rates, $K(\gamma)$ decays monotonically from a maximum initial value $K_i$ to a strain-independent plateau $K_f$ with a much lower stiffness value, indicating the 'soft' regime. The two 0.5 mg/ml DNA systems (Fig 4a,c) show similar steady monotonic increases in stiffness with increasing $\dot{\gamma}$, with 0.5:0.7 exhibiting ~2-fold larger values for each strain rate. This trend, which can also be seen in Fig 4d in which we plot $K_i$ vs $\dot{\gamma}$ for all three systems, corroborates the physical picture that the resistance exerted by both 0.5 mg/ml systems in the nonlinear regime arises primarily from polymer buildup rather than entropic stretching. As $\dot{\gamma}$ increases, the DNA cannot as easily disentangle (a dissipative process) so the elastic contribution to the nonlinear force response (measured by $K$) is stronger. The increased stiffness for the 0.5:0.7 system compared to 0.5:0 can be understood simply as the contribution of the stiffer microtubules that are unable to undergo dissipative processes on the timescale of the strain.

Notably, the 0.65:0 blend exhibits starkly different behavior, including much weaker and non-monotonic dependence of $K(\gamma)$ on strain rate as well lower stiffness values than either of the 0.5 mg/ml systems in the initial 'stiff' regime (Fig 4b,d). This effect clearly indicates that different physical phenomena underlie the nonlinear stress response for 0.65:0 compared to 0.5:0 and 0.5:0.7, as our Fig 3 results indicate. Reduced stiffness and lack of increase with $\dot{\gamma}$ are in line with enhanced DNA flow alignment and shear thinning, which we postulate is facilitated by ring-linear threading [18,20,26,29,49,65]. Namely, polymers become increasingly stretched and aligned along the strain path as $\dot{\gamma}$ increases, which reduces the entanglement density and thus stiffness.

However, as seen in the insets of Figs 4a-c, these trends do not hold in the 'soft' regime in which all systems display lower strain-independent $K$ values that increase monotonically with $\dot{\gamma}$, suggesting that different mechanisms dictate the different regimes. As most clearly seen in Fig 4e, which shows the strain-averaged stiffness values in this final plateau regime $K_f$, the 0.65:0 blend tracks closely with the 0.5:0.7 composite while the 0.5:0 blend shows lower stiffness values and weaker $\dot{\gamma}$ dependence. This result corroborates that the force response in this softer regime is dominated by dissipative processes such as disentanglement, de-threading and reorienting, and that the system with the weakest constraints (0.5:0) is most easily able to dissipate stress. Conversely, the 0.65:0 and 0.5:0.7 retain more stiffness due to the long-lived threading constraints and the stiffness of the microtubules, respectively, both of which hinder dissipative relaxation on the timescales of the nonlinear strains.

*Nonlinear Relaxation Dynamics*

Our results shown in Figs 3 and 4 suggest varying relaxation modes and timescales for the different systems and strain rates. To directly measure the relaxation dynamics, we record the relaxation of the force exerted by the polymers on the probe as it is held fixed at the final strain position (Fig 1d). Figs 5a, S2, and S3 shows that all systems maintain a non-zero residual force $F_R$ at the end of the relaxation phase, suggestive of very slow relaxation modes. The 0.5:0.7 composite maintains the largest $F_R$ across all strain rates, ~5-fold larger than the 0.5:0 system, owing to the very slow relaxation modes of the stiff microtubules (Fig 5b). Further, $F_R$ values for both 0.5 mg/ml DNA systems display minimal dependence on $\dot{\gamma}$, in contrast to 0.65:0 in which the residual force steadily decreases from values comparable to 0.5:0.7 to those that more closely align with 0.5:0. This trend is similar to that of the maximum force $F_M$ and effective viscosity $\eta_{en}$ during strain (Fig 3e) in which both 0.5 mg/ml systems exhibit similar $\dot{\gamma}$ dependence, while $F_M$ and $\eta_{en}$ values for 0.65:0 transition from being comparable to 0.5:0.7 to more closely matching 0.5:0. At lower strain rates, threadings and entanglements in the 0.65:0 system provide strong constraints with slow relaxation modes, similar to those of stiff microtubules. However, with increasing $\dot{\gamma}$ the DNA increasingly aligns with the strain to reduce the effective entanglement density in the vicinity of the strain. Contributions from strain-induced disentanglement and dethreading are also likely more prevalent at faster strain rates.

While $F_R$ is a measure of the force that remains elastically stored in the system, it does not unequivocally indicate the degree to which each system sustains or relaxes strain-induced force $F_M$ because $F_M$ also depends on the system and strain rate. To quantify the relative elastic storage of each system, we compute the fractional force sustained during the relaxation phase, $f_R = F_R/F_M$, which varies from $f_R = 0$ for purely viscous systems to $f_R = 1$ for elastic systems (Fig 5c). As shown, $f_R$ decreases with increasing $\dot{\gamma}$ for all three networks, with 0.65:0 exhibiting modestly stronger decrease at higher $\dot{\gamma}$, in line with the system-dependent nonlinear viscosity thinning we observe (Fig 3e). Moreover, $f_R$ for 0.65:0 is indistinguishable from 0.5:0.7 for all but the fastest strain rate, demonstrate the strong constraints and suppressed dissipative modes that threadings (0.65:0) and rigid MTs (0.5:0.7) confer. The 0.5:0 blend, which has fewer entanglements and threadings and no stiff MTs, exhibits ~2× smaller $f_R$ values across all strain rates, indicative of the most dissipative (viscous) relaxation dynamics.

To further elucidate the relaxation modes that give rise to these observed trends, we fit each force relaxation curve to a sum of three exponential decays, $F(t) = F_R + C_1 e^{-t/\tau_1} + C_2 e^{-t/\tau_2} + C_3 e^{-t/\tau_3}$, with characteristic timescales $\tau_1$, $\tau_2$ and $\tau_3$ and non-zero residual $F_R$ (Fig 5b, green dashed lines). We have previously shown the relaxation dynamics of entangled ring-linear DNA blends and flexible-stiff polymer composites can be well-described by this function[18,32,66]. Each fit yields three well-separated relaxation timescales $\tau_i$ (Fig 5d) that depend on $\dot{\gamma}$ and network composition. All three $\tau_i$ values generally decrease with increasing $\dot{\gamma}$ for all systems, with the magnitudes of each $\tau_i$ being generally largest (slowest relaxation) for 0.65:0 compared to the two 0.5 mg/ml DNA systems. Interestingly, 0.5:0 and 0.5:0.7 exhibit similar values for all three time constants $\tau_i$, despite the large difference in $F_R$ and $f_R$ between the two systems, suggesting that DNA dynamics dictate the relaxation dynamics of the DNA-MT composite while the relaxation modes of the MTs are frozen out at these nonlinear strain rates.

Additionally, all relaxation timescales are shorter than the those measured in the linear regime, indicating strain-induced disentanglement and de-threading. However, $\tau_1$ values are comparable to $t_{soft}$ (Fig 3f) as well as the predicted entanglement times for linear DNA ($\tau_{e,0.5} \simeq 0.17$ s, $\tau_{e,0.65} \simeq 0.10$ s). Taken together, we expect $\tau_1$ to be a measure of the entanglement timescale in the nonlinear regime, which dictates the transition to the softer regime of the force response ($\tau_1 \simeq t_{soft}$). The agreement of $\tau_1$ with $\tau_e$ for comparable linear DNA systems is an indicator of forced de-threading, whereby linear chains are pulled along with the moving probe while rings, which are less easily able to stretch and align with the strain, are de-threaded and diffuse into the wake behind the probe. Within this physical picture, the network that the probe measures is composed primarily of linear chains which are denser in the 0.5:0 systems compared to 0.65:0, due to enhanced polymer build-up, resulting in faster/lower $\tau_1$ values compared to $\tau_{1,0.65}$. Similarly, the longest measured relaxation timescales $\tau_3$ are on the order of the predicted disengagement times for linear DNA ($\tau_{D,0.5} \simeq 7.4$ s, $\tau_{D,0.65} \simeq 10$ s), which are substantially shorter than those measured for the blends in the linear regime. This finding corroborates the described mechanisms of (1) forced de-threading, which minimizes the contribution of ring DNA to the nonlinear response, and (2) forced disentanglement, which lowers the effective entanglement density and thus lowers $\tau_D$. The intermediate timescale $\tau_2$ is likely a result of elastic stretching and retraction similar to Rouse modes[67,68]. Indeed, several previous studies on entangled DNA have shown agreement between experimentally measured $\tau_2$ values and predicted Rouse times $\tau_R$ [18,69,70]. Insofar as the physical picture described above is correct, we expect $\tau_2$ to match more closely with $\tau_R$ for linear DNA ($\tau_{R,L} \simeq 0.51$ s versus ring DNA ($\tau_{R,L} \simeq 0.17$ s), which is indeed what we observe (Fig 5d).

While our measured time constants are comparable to predicted relaxation timescales for all systems, as described above, they generally decrease with $\dot{\gamma}$ in a system-dependent manner. To more closely examine this rate dependence, we normalize $\tau_1, \tau_2$ and $\tau_3$ by their respective values at the lowest strain rate ($\dot{\gamma} = 9.4$ s$^{-1}$) (Fig 5e). In all cases, the 0.5:0.7 composite exhibits the weakest $\dot{\gamma}$ dependence, an indicator of the enhanced elasticity that the MTs provide that suppresses viscous rate dependence. Likewise, 0.5:0, the most dissipative system, exhibits the strongest $\dot{\gamma}$ dependence. Interestingly, $\tau_i/\tau_{i,9.4}$ for 0.65:0 tracks closely with that for 0.5:0 for the two faster timescales ($\tau_1, \tau_2$), while it tracks with 0.5:0.7 for $\tau_3$. This result aligns with Fig 5b, in which the nonlinear force response of 0.65:0 matches with 0.5:0.7 at slow strain rates (long timescales) then transitions to agreement with 0.5:0 for fast strain rates (short timescales). Moreover, as we describe above, $\tau_1$ and $\tau_2$, which we understand to equate to $\tau_e$ and $\tau_R$, are dictated primarily by the dynamics and properties of the DNA whereas $\tau_3$ ($\simeq \tau_D$) is controlled by the system constraints (i.e., entanglements and threadings) which are comparable in strength for 0.65:0 and 0.5:0.7, as described above.

Finally, we turn to evaluating the contribution of each relaxation mode, corresponding to $\tau_1, \tau_2$ and $\tau_3$, to the overall relaxation of each system. We quantify this contribution by computing the fractional coefficients $\phi_i = C_i/\sum C_i$ associated with each decay mode (Fig 5f). We observe that the coefficients associated with $\tau_1$ and $\tau_3$ ($\phi_1$ and $\phi_3$) increase and decrease, respectively, with increasing $\dot{\gamma}$, which we postulate arises from the flow alignment, disentanglement and de-threading that occurs at higher strain rates. Further corroborating this interpretation is the fact that both the magnitude and $\dot{\gamma}$ dependence of $\phi_3$ are lowest for 0.5:0, which is the system with the fewest and weakest constraints. Interestingly, $\phi_2$ exhibits much weaker $\dot{\gamma}$ dependence for all systems, which aligns with our interpretation of $\tau_2$ equating to the Rouse time $\tau_R$ which is relatively insensitive to entanglement density and timescales. We do note, however, that there is a weak non-monotonic trend, most pronounced for 0.65:0, whereby $\phi_2$ peaks at $\dot{\gamma} \approx 42$ s$^{-1}$. Insofar as $\tau_2 \approx \tau_R$ this peak indicates maximal elastic stretching and retraction, in line with our description above, in which DNA chains are increasingly stretched along the strain path as $\dot{\gamma}$ increases to a resonant rate, after which the strain is too fast for the DNA to efficiently couple to it, such that stretching is reduced.

**Conclusions**

Here, we present comprehensive linear and nonlinear microrheology measurements of entangled blends of ring and linear DNA and their composites with microtubules. We judiciously chose these systems to shed light on the highly debated roles that polymer threadings and entanglements play in the rheological properties of ring-linear polymer blends, and the extent to which these synergistic topological interactions can mediate the rheology of composites of flexible and stiff polymers. Moreover, we sought to elucidate the intriguing emergent mechanical properties that mixtures of polymers of varying topologies and stiffnesses have been shown to exhibit [32,55,71], many of which cannot be predicted from the properties of the constituents or reproduced in single-constituent systems.

As previous studies have demonstrated that threading of rings by linear chains is most pervasive in entangled blends with comparable concentrations of rings and linear chains [18,28], we focus our studies on blends with equal mass fractions of ring and linear DNA. Moreover, the surprisingly large effect that DNA topology (ring vs linear) has one the emergent rheological properties of DNA-MT composites [32] and other mixed polymer systems [53], motivate our incorporation of microtubules into ring-linear blends. Finally, we purposefully design the three networks we study here, defined by the concentrations of DNA and tubulin ($c_D:c_T = $ 0.5:0, 0.65:0, 0.5:0.7), to have relatively modest differences in their composition, with a 30% difference in concentration between the two DNA blends (0.5:0, 0.65:0) and similar mesh sizes for the DNA-MT composite (0.5:0.7) and concentrated ring-linear blend (0.65:0). In this way, we can confidently attribute the robust variations in rheological properties we observe to modulations in the interactions between the polymers rather than largescale structural differences.

We find that in the linear viscoelastic regime, relaxation modes are dictated by the presence of rings and threading events which prolong fast and slow relaxation timescales, respectively. However, microtubules are required to substantially augment the degree of elasticity and shear thinning of blends, with both effects becoming frozen out at high frequencies due to the slow intrinsic relaxation modes of the microtubules. Moreover, while linear rheological properties of the DNA-MT composite display markedly different frequency-dependences compared to the DNA blends, in the nonlinear regime it is the higher concentration DNA blend (0.65:0) that displays a rate-dependence that is distinct from that of the two 0.5 mg/ml systems. The distinct nonlinear features of 0.65:0 include enhanced shear thinning, slowed relaxation timescales, reduced stiffness that displays a non-monotonic rate dependence, and dynamics that transition from being akin to 0.5:0 at the lowest rates to 0.5:0.7 at the fastest rates. Our collective results suggest that these emergent rheological properties arise from the coupled effects of strain-induced flow alignment, de-threading, disentanglement, and build-up at the leading edge of the moving probe – all of which contribute to varying degrees and over different timescales in the various networks.

Our results are broadly applicable to understanding the rheological properties of diverse entangled polymeric systems, in particular those that include 'endless' ring polymers, polymers of varying stiffnesses, and mixtures thereof. Such materials, which continue to fascinate and frustrate polymer scientists, are also ubiquitous in biology (e.g., cells, nucleus, cytoskeleton, mucus, cartilage) and industrial applications (e.g., flow regulation, super absorption, miscibility, adhesion), making understanding their mechanical properties of broad interest and importance.

**Supplementary Materials**

Section S1: Expanded descriptions of predicted and measured length and time scales for entangled polymers. Figure S1. Nonlinear force response of ring-linear DNA blends (0.5:0, 0.65:0) and DNA-MT composite (0.5:0.7); Figure S2. Relaxation of strain-induced force in ring-linear DNA blends (0.5:0, top; 0.65:0, middle) and DNA-MT composite (0.5:0.7, bottom) following nonlinear strains; Figure S3. Fits of force relaxation curves to a sum of exponential decays.

**Acknowledgements**

We acknowledge funding from the Air Force Office of Scientific Research awards (FA9550-17-1-0249, FA9550-21-1-0361) to RMRA.

**Conflicts of Interest**

The authors have no conflicts to disclose.

**Figures**

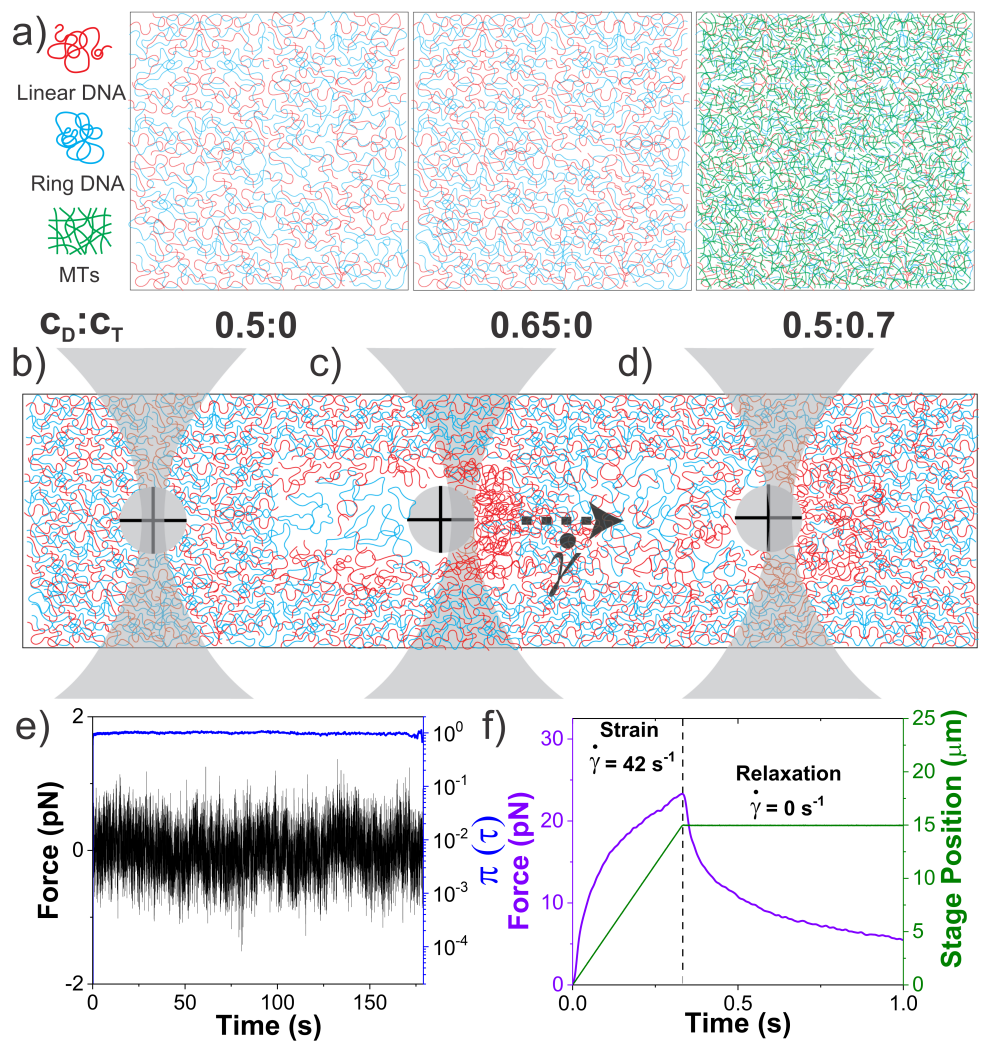

**Figure 1. Measuring the linear and nonlinear microrheological properties of entangled ring-linear DNA blends and their composites with rigid microtubules.** (a) Cartoons of the ring-linear DNA blends and DNA-MT composite, defined by the mass concentrations (mg/ml) of DNA $c_D$ (0.5 or 0.65) and tubulin $c_T$ (0 or 0.7). (b) Cartoon of microsphere (grey circle) with radius $R = 2.25$ μm, embedded in a ring-linear DNA blend and trapped using a focused Gaussian laser beam (grey). Linear microrheology measurements are performed by measuring the thermal deviations of the bead from the trap center in equilibrium. (c) For nonlinear microrheology measurements, the same optically trapped bead is displaced 15 μm ($\gamma = 3$) through the blend at logarithmically spaced speeds $v = 10$–$200$ μm/s, corresponding to strain rates $\dot{\gamma} = 3v/\sqrt{2}R = 9.4 - 189$ s$^{-1}$. (d) The bead motion is then halted and the surrounding polymers are allowed to relax back to equilibrium. (e) Sample thermal oscillation data (black), captured for 200 seconds at 20 kHz, for the 0.5:0 DNA blend. From the thermal oscillations, we compute normalized mean square displacements $\pi(\tau)$ (blue) which we use to extract viscoelastic moduli using GSER (see Methods). (f) For nonlinear microrheology measurements we record the stage position (green) and force exerted on the trapped bead (violet) during (0.075-1.5 s), and following (15 s) the bead displacement (delineated by dashed lines) at 20 kHz. Data shown is for the 0.5:0 DNA blend at $\dot{\gamma} = 42$ s$^{-1}$.

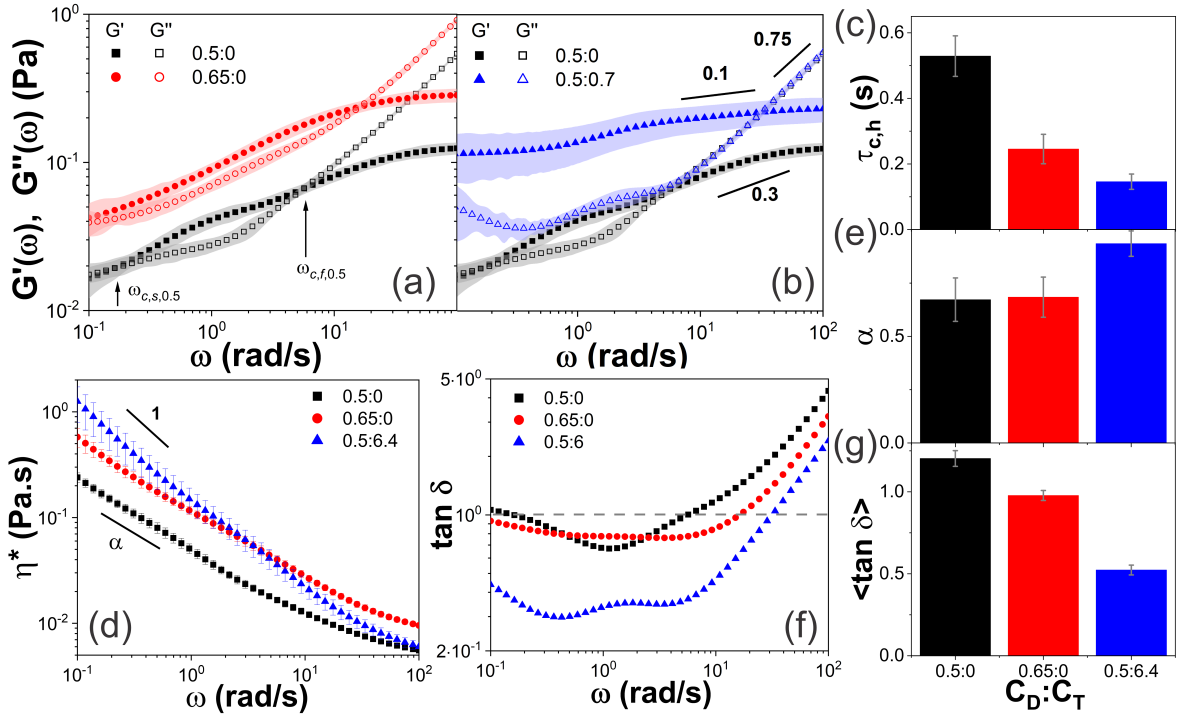

**Figure 2. Linear viscoelasticity of ring-linear DNA blends and DNA-MT composites exhibit complex dependence on polymer concentration and stiffness.** (a,b) Frequency-dependent elastic and viscous moduli, $G'(\omega)$ (closed symbols) and $G''(\omega)$ (open symbols) of a ring-linear DNA blend with $c_D = 0.5$ mg/ml (black) compared to (a) a higher concentration ring-linear blend ($c_D = 0.65$ mg/ml, red) and (b) its composite with microtubules polymerized from tubulin dimers of concentration $c_T = 0.7$ mg/ml (blue). The shaded region surrounding each curve represents the standard error across 15 trials. The legend demarcates each network by the ratio of DNA to tubulin concentrations $c_D:c_T$. Arrows point to the low (slow) and high (fast) crossover frequencies $\omega_{c,s}$ and $\omega_{c,f}$ for 0.5:0, which indicate the slowest/fastest characteristic relaxation timescales $\tau_{c,s/f} = 2\pi/\omega_{c,s/f}$ of the blend. (c) The fast relaxation timescale $\tau_{c,s}$ for 0.5:0 (black), 0.65:0 (red) and 0.5:0.7 (blue), determined from the corresponding $\omega_{c,f}$ indicates the entanglement time $\tau_e$. (d) Complex viscosity $\eta^*(\omega)$, determined from $G'(\omega)$ and $G''(\omega)$ curves shown in (a) and (b), exhibit shear-thinning $\eta^*(\omega) \sim \omega^{-\alpha}$ with (e) exponents $\alpha$, computed from power-law fits to $\eta^*(0.1 \text{ rad/s} < \omega < 1 \text{ rad/s})$. (f) Loss tangent $\tan\delta = G''(\omega)/G'(\omega)$ versus $\omega$ computed from data shown in (a) and (b) with a dashed grey line representing $G''(\omega) = G'(\omega)$ ($\tan\delta = 1$). The frequencies at which each curve crosses the dashed line correlate with $\omega_{c,s/f}$. (g) Frequency-averaged loss tangents $\langle\tan\delta\rangle$ for all three systems indicate the extent to which each system exhibits elastic versus viscous dynamics. $\langle\tan\delta\rangle > 1$ indicates that dissipative mechanisms dictate dynamics while $\langle\tan\delta\rangle < 1$ indicates more elastic-like dynamics with lower values indicating relatively more elasticity (less dissipation).

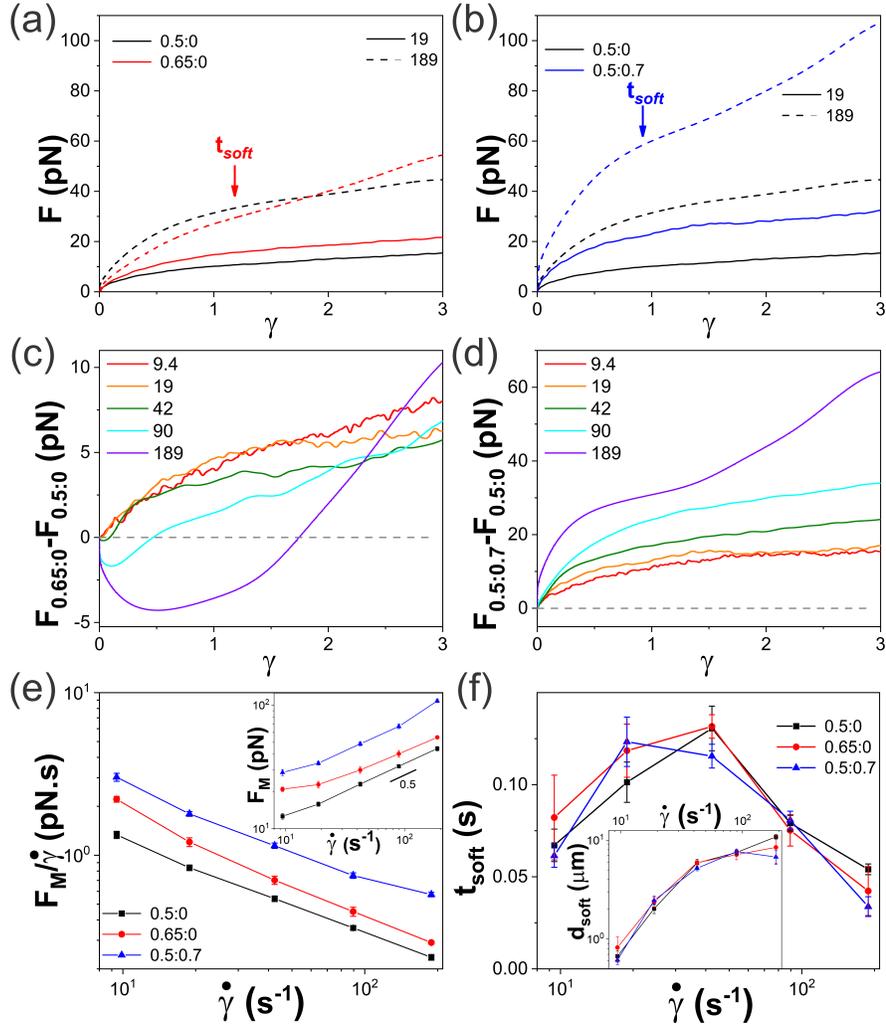

**Figure 3. The nonlinear force response reveals unique nonlinear viscoelasticity for highly entangled DNA blends (0.65:0) dictated by ring-linear threadings.** (a-b) Measured force $F$ in response to nonlinear strain with rates of $\dot{\gamma} = 19$ s$^{-1}$ (solid) and $189$ s$^{-1}$ (dashed), as denoted in the legend, for a ring-linear DNA blend with $c_D = 0.5$ mg/ml (0.5:0, black) compared to (a) a higher concentration ring-linear blend ($c_D = 0.65$ mg/ml, red) and (b) its composite with microtubules polymerized from tubulin dimers of concentration $c_T = 0.7$ mg/ml (blue). Data for all strain rates are shown in Fig S1. (c-d) The force response of the 0.5:0 blend, $F_{0.5:0}$, subtracted from the corresponding force curves for (c) 0.65:0 ($F_{0.65:0}$) and (d) 0.5:0.7 ($F_{0.65:0}$) for all strain rates $\dot{\gamma}$ (listed in the legend in s$^{-1}$) show markedly different effects of adding more DNA (c) versus MTs (d) to the 0.5:0 DNA blend. (e) The effective nonlinear viscosity $\eta_{en}(\dot{\gamma})$, computed by dividing the maximum force $F_M$ reached at the end of the strain (see inset) by the strain rate $\dot{\gamma}$, for 0.5:0 (black), 0.65:0 (red) and 0.5:0.7 (blue) demonstrates nonlinear shear thinning behavior that is distinct from the linear regime. Black line in inset, representing power-law scaling of $F_M \sim \dot{\gamma}^{0.5}$, indicates viscoelastic behavior in between purely elastic ($F_M \sim \dot{\gamma}^0$) and viscous ($F_M \sim \dot{\gamma}^1$) scaling. (f) The softening timescale $t_{soft}$, indicated by arrows in (a) and (b), and lengthscale $d_{soft}$ (inset), quantify the time and lengthscales at which the force response transitions from the initial stiffer regime to the eventual softer regime respectively. $t_{soft}$ and $d_{soft}$ are plotted versus $\dot{\gamma}$ for all three systems as indicated in the legend.

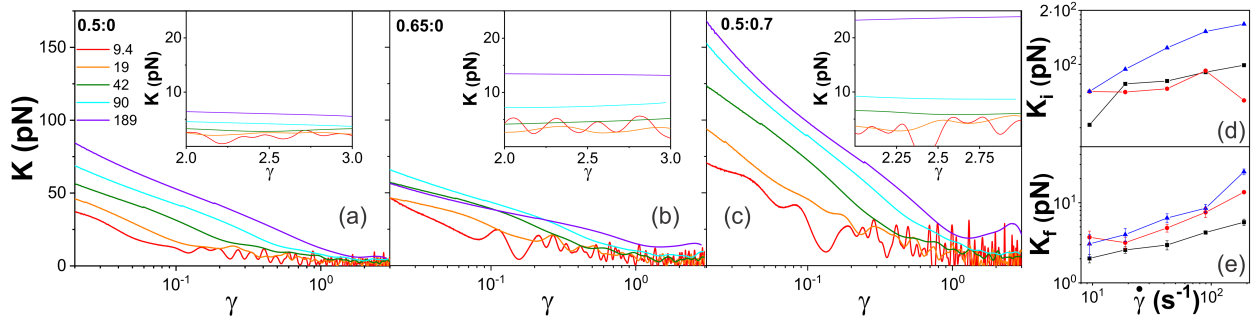

**Figure 4. The stiffness of ring-linear polymer blends and composites exhibits surprising dependence on network composition and strain rate.** (a-c) The effective differential modulus of stiffness $K(\gamma) = dF/d\gamma$, a measure of how elastic-like or stiff each system is, plotted versus strain $\gamma$ for strain rates $\dot{\gamma}$ (s$^{-1}$) listed in the legend for (a) 0.5:0, (b) 0.65:0 and (c) 0.5:0.7. $K(\gamma)$ for all three systems decreases from a maximum initial value $K_i$ (plotted in d) to a non-zero plateau, amplified in the corresponding inset, with a strain-averaged value $K_f$ (plotted in e). We denote the region over which $K$ decreases as the 'stiff' regime and the terminal plateau as the 'soft' regime. (d) While the initial stiffness $K_i$ steadily increases with increasing strain rate $\dot{\gamma}$ for 0.5:0 (black) and 0.5:0.7 (blue), it exhibits a non-monotonic dependence for 0.65:0 (red). (e) The strain-averaged stiffness in the soft regime $K_f$ generally increases with increasing $\dot{\gamma}$ for all systems, but the magnitude and degree of increase depends on the system composition (0.5:0 (black), 0.65:0 (red), or 0.5:0.7 (blue)).

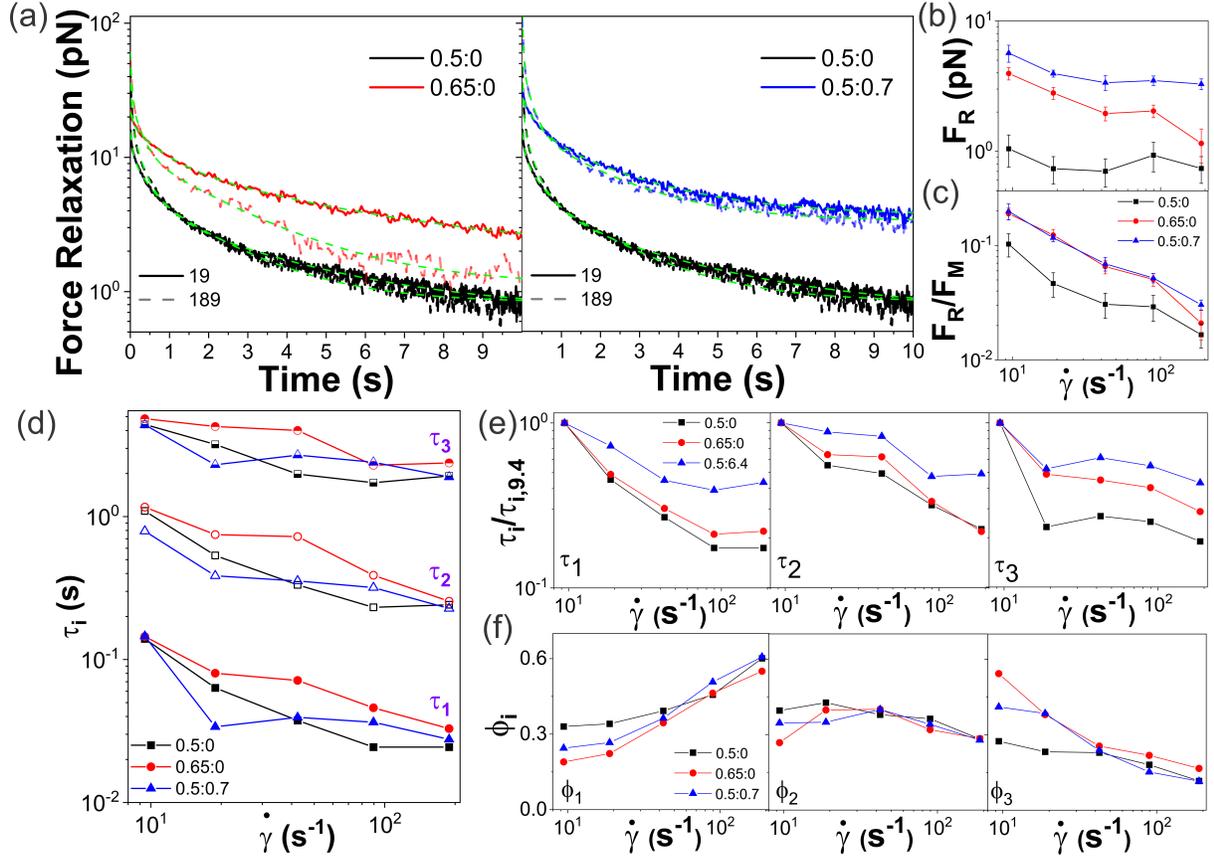

**Figure 5. Force relaxation following nonlinear straining exhibits multi-mode relaxation dictated by strain-induced de-threading of rings and polymer stiffness.** (a) Force relaxation of 0.5:0 (black), 0.65:0 (red, left panel) and 0.5:0.7 (blue, right panel) as a function of time following strain for $\dot{\gamma} = 19$ s$^{-1}$ (solid) and 189 s$^{-1}$ (dashed). Data for all strain rates are shown in Fig. S2. Each relaxation curve is fit to a sum of three exponential decays and a residual, $(t) = F_R + C_1 e^{-t/\tau_1} + C_2 e^{-t/\tau_2} + C_3 e^{-t/\tau_3}$, indicated by the green dashed lines, all of which have adjusted $R$-squared values of ≥0.99. Fits for all strain rates are shown in Fig S3. (b) All networks exhibit sustained elastic memory following strains of all rates $\dot{\gamma}$, as shown by the non-zero residual force $F_R$ measured at the end of the relaxation phase, which is highest when MTs are present (0.5:0.7, blue) and lowest for 0.5:0 (black). (c) Data shown in (b), normalized by the corresponding maximum force $F_M$ (see Fig 3e inset) indicates the degree to which each network elastically stores versus viscously dissipates strain-induced stress after nonlinear straining. The addition of either DNA (0.65:0, red) or MTs (0.5:0.7, blue) to 0.5:0 (black) similarly increases the elastic storage across all $\dot{\gamma}$. (d) Characteristic relaxation time constants $\tau_1$ (filled), $\tau_3$ (open), and $\tau_3$ (half-filled), determined from fits (Fig S3) and plotted against $\dot{\gamma}$, show that all relaxation timescales $\tau_i$ generally decrease with increasing $\dot{\gamma}$ for all systems [0.5:0 (black), 0.65:0 (red), 0.5:0.7 (blue)]. (e) $\tau_1$ (left), $\tau_2$ (middle), and $\tau_3$ (right), normalized by their corresponding values for the lowest strain rate $\dot{\gamma} = 9.4$ $s^{-1}$ ($\tau_{i,9.4}$) are plotted against $\dot{\gamma}$ for all three systems. (f) Fractional coefficients $\phi_i = C_i/(C_1 + C_2 + C_3)$ associated with each timescale $\tau_i$ shown in the panel above in (e) are plotted against $\dot{\gamma}$. For reference, $\phi_1 + \phi_2 + \phi_3 = 1$ and larger values $\phi_i$ values indicate larger relative contributions to the stress relaxation.

**Polymer threadings and rigidity dictate the viscoelasticity and nonlinear relaxation dynamics of entangled ring-linear blends and their composites with rigid rod microtubules**

Karthik R. Peddireddy, Ryan Clairmont, and Rae M. Robertson-Anderson*

*Department of Physics and Biophysics, University of San Diego, 5998 Alcala Park, San Diego, CA 92110, United States*

**Physical properties of DNA solutions:** Contour lengths of the ring and linear DNA are fixed at $L = 115$ kbp (~39 μm), but due to their end-closure, rings have ~$1.6x$ smaller radius of gyration $R_G$ than their linear counterparts[1]. Using the previously reported values of $R_{G,R} \simeq 541$ nm and $R_{G,L} \simeq 885$ nm for ring and linear 115 kbp DNA[1], we determine the polymer overlap concentration of the ring-linear DNA blend via $c^*_{RL} = (3/4\pi)(M/N_A)/(f_L R^3_{G,L} + f_R R^3_{G,R})$ where $M$ is the DNA molecular weight[2]. Our computed value of $c^*_{RL} \simeq 710$ μg/mL yields $c_{D,\downarrow} = 0.5$ mg/ml $\simeq 7c^*_{RL}$ and $c_{D,\uparrow} = 0.65$ mg/ml $\simeq 9c^*_{RL}$. We previously showed that the critical entanglement concentration for both linear and ring DNA is $c_e \simeq 6c^*_L$.[3] Diffusion coefficients of the linear and ring 115 kbp DNA in dilute conditions are $D_L = 0.3$ μm²/s and $D_R = 0.42$ μm²/s, respectively[1]. The corresponding Rouse time $\tau_R = 2R_G^2/\pi^2 D$, which is the time over which elastic relaxation of a stretched polymer occurs, is $\tau_{R,L} \simeq 0.51$ s and $\tau_{R,R} \simeq 0.14$ s for linear and ring 115 kbp DNA.

**Theoretical lengthscales and timescales**: The Doi-Edwards (reptation) model[2] predicts the dynamics of entangled linear polymers[2]. With the DE framework, the fastest relaxation timescale is the entanglement time $\tau_e = a^4/24R_G^2 D$, which is the time scale over which thermally diffusing chain segments reach the edge of the reptation tube of diameter $a$. The slowest relaxation timescale is the disengagement time $\tau_D = 36R_G^4/\pi^2 a^2 D$, which is the time over which the polymer reptates out of its initial deformed tube. The tube diameter is computed using the relation $a = (24N_e/5)^{1/2} R_G$ where $N_e = (4/5)cRT/MG_N^0$ is the number of entanglements per chain, $M$ is the polymer molecular weight and $G_N^0$ is the plateau modulus. The corresponding entanglement length is $L_e = L/N_e$. Combining the predicted scaling $G_N^0 \sim c^2$ with the expressions given above yields the following scaling relations for the fast and slow relaxation timescales: $\tau_e \sim a^4 \sim L_e^2 \sim c^{-2}$ and $\tau_D \sim L_e^{-1} \sim c$. To estimate the corresponding length and time scales for the ring DNA, we use the pom-pom ring model[4] which predicts the tube diameter for rings via the relation $a_R/a_L = (5N^{-0.4})^{1/2}$, and the disengagement time for rings from the relation $\tau_{D,R}/\tau_{D,L} = (a_R/a_L)^2 N^{-0.4}$.

Using our previously reported value of $G_N^0 \simeq 0.2$ Pa for linear 115 kbp DNA at $c = 1$ mg/ml[5], the scaling $G_N^0 \sim c^2$ [2], and the reported physical values of the DNA[1], we estimate the theoretical length and time scales for 0.5 mg/ml and 0.65 mg/ml entangled linear and ring DNA solutions which are tabulated below.

| DNA Topology | $c$ (mg/ml) | $L_e$ (μm) | $a$ (μm) | $\tau_e$ (s) | $\tau_R$ (s) | $\tau_D$ (s) |
|---|---|---|---|---|---|---|
| Linear | 0.5 | 10.2 | 0.99 | 0.17 | 0.51 | 7.4 |
| Linear | 0.65 | 7.8 | 0.87 | 0.10 | 0.51 | 9.6 |
| Ring | 0.5 | - | 0.67 | 0.04 | 0.14 | 0.32 |
| Ring | 0.65 | - | 0.59 | 0.02 | 0.14 | 0.41 |

We further estimate the effective tube diameter for the ring-linear DNA blends from the tube diameters of the pure ring and linear polymer solutions. We estimate $a_L$ for the $f_L \simeq 0.5$ fraction of linear DNA in the 0.5:0 and 0.65:0 blends (i.e., ~0.25 mg/ml and ~0.375 mg/ml) as $a_L \simeq 1.4$ µm and ~1.2 µm, respectively. The corresponding tube radius $a_R$ values for the $f_R \simeq 0.5$ ring fraction are ~0.95 µm and ~0.84 µm for $c_D = 0.5$ mg/ml and 0.65 mg/ml, respectively. Finally, to estimate an effective tube diameter $d_T = 2a$ for each $f_L \simeq f_R \simeq 0.5$ blend, we use the relation $d_T^{-3} = [(2a_L)^{-3} + (2a_R)^{-3}]$ that considers the density of each cubic tube diameter $d_T^3$ to arrive at values of $d_T \simeq 1.74$ µm and ~1.53 µm for the 0.5:0 and 0.65:0 blends, respectively[6]. The theoretical mesh size of the entangled MT network for 0.7 mg/ml tubulin is $\xi = 0.89/c_T^{1/2} \simeq 1.1$ µm which is comparable to the tube diameter of the 0.65:0 blend[7].

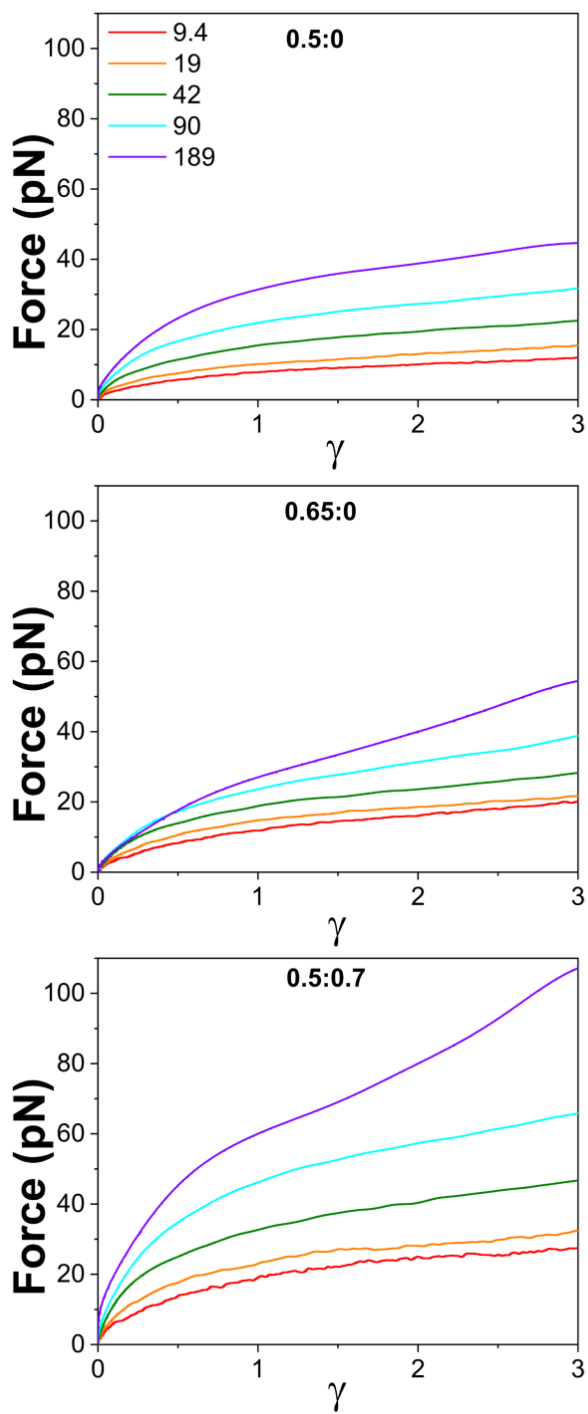

**Figure S1. Nonlinear force response of ring-linear DNA blends (0.5:0, 0.65:0) and DNA-MT composite (0.5:0.7).** Measured force in response to 15 μm ($\gamma = 3$) strains with $\dot{\gamma}$ listed in s$^{-1}$ in legend (top left corner of the top figure). Stage displacement data $x$ is converted to strain via $\gamma = x/2R$ where $R = 2.25$ μm is the bead radius. The concentrations of DNA ($c_D$) and tubulin ($c_T$) in mg/ml are listed at the top of each plot.

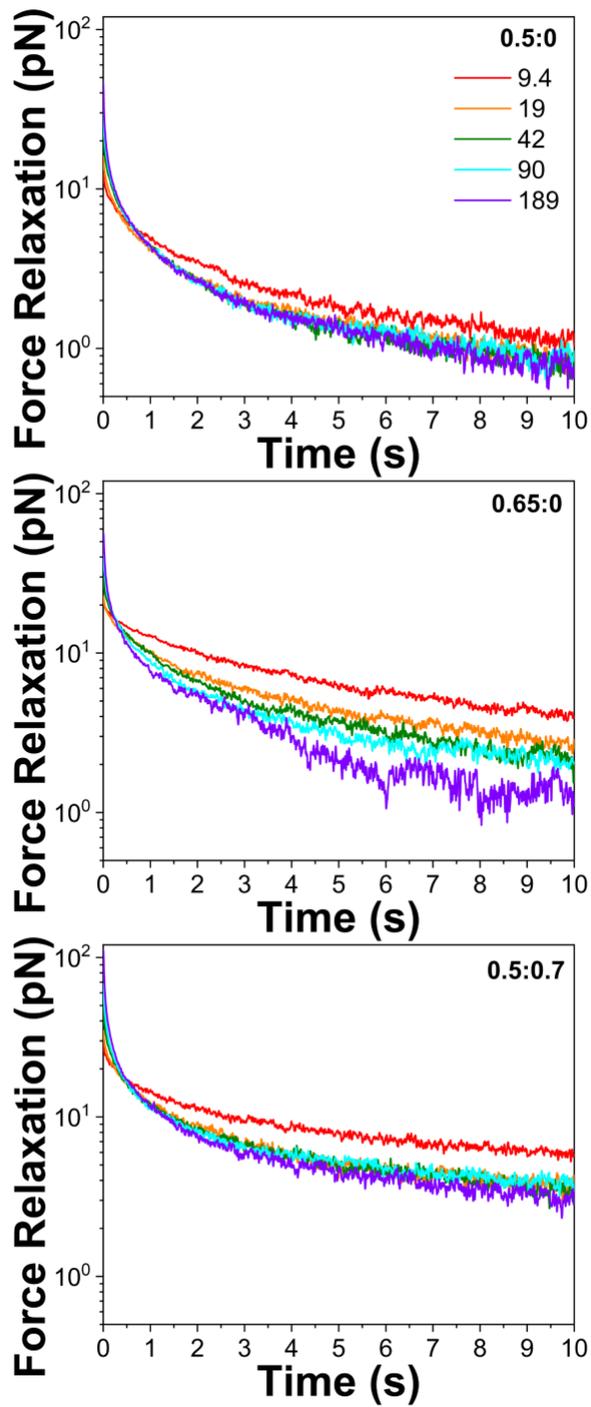

**Figure S2. Relaxation of strain-induced force in ring-linear DNA blends (0.5:0, top; 0.65:0, middle) and DNA-MT composite (0.5:0.7, bottom) following nonlinear strains.** Measured relaxation of force following 15 μm strains with $\dot{\gamma}$ listed in units of s$^{-1}$ in legend (top plot). The concentrations of DNA ($c_D$) and tubulin ($c_T$) are listed in mg/ml at the top right corner of each plot.

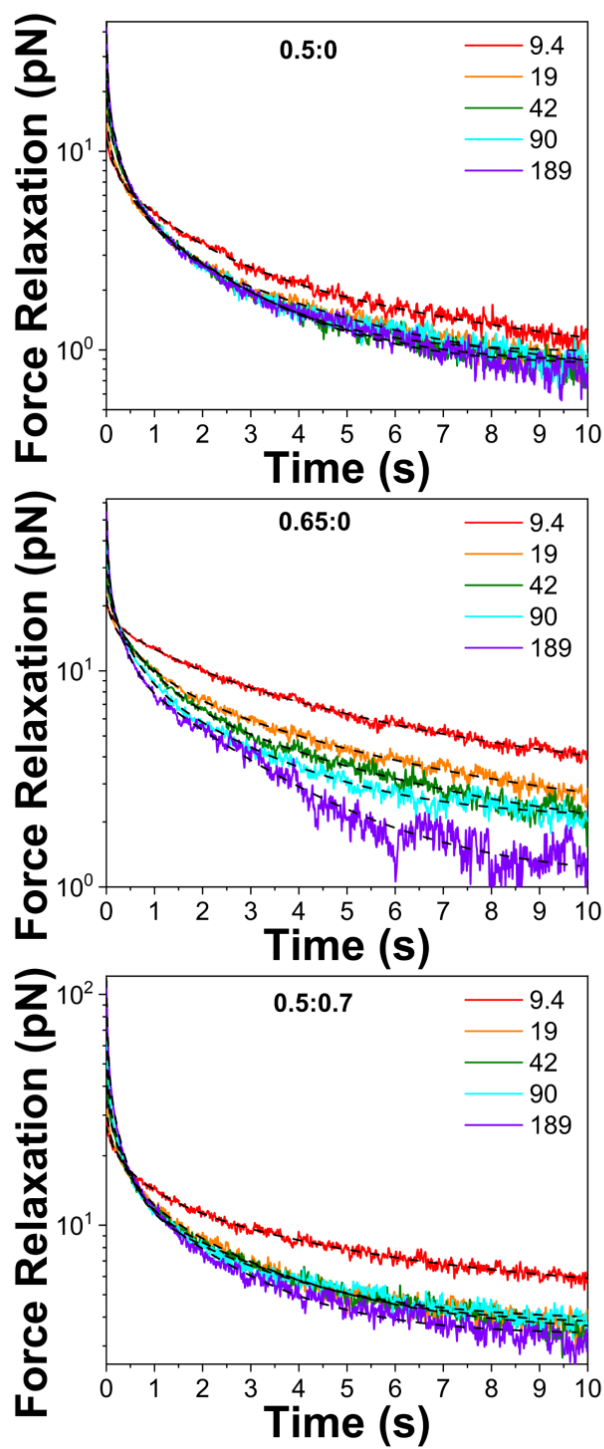

**Figure S3. Fits of force relaxation curves to a sum of exponential decays.** All force relaxation curves (also shown in Fig S2) are well-fit to a sum of three exponential decays and a residual force $F_R$: $F(t) = F_R + C_1 e^{-t/\tau_1} + C_2 e^{-t/\tau_2} + C_3 e^{-t/\tau_3}$. Fits are shown as black dashed lines and all have adjusted $R$-squared values of $\geq 0.99$. Force relaxations following strains with varying strain rates $\dot{\gamma}$, listed in the legends in $s^{-1}$, are shown for 0.5:0 (top), 0.65:0 (middle) and 0.5:0.7 (bottom).

**Supplementary References**